\documentclass[conference]{IEEEtran}
\IEEEoverridecommandlockouts

\usepackage{amssymb}

\ifCLASSINFOpdf
   \usepackage[pdftex]{graphicx}
   \graphicspath{{../pdf/}{../jpeg/}}
  \DeclareGraphicsExtensions{.pdf,.jpeg,.png}
\else
\fi
\usepackage[cmex10]{amsmath}
\usepackage{txfonts}
\usepackage{url}

\begin{document}
%
\title{Agent-based modelling of social organisations\thanks{The research leading to these results has received funding from the European Community's Seventh Framework Program (FP7/2007-2013) under grant agreement n° 218086.}}

\author{\IEEEauthorblockN{Jaros\l aw Ko\'zlak and Anna Zygmunt}
\IEEEauthorblockA{
Department of Computer Science\\
AGH University of Science and Technology\\
Al. Mickiewicza 30, 30-059 Kraków, Poland\\
\{kozlak, azygmunt\}@agh.edu.pl
}
}


%


\maketitle

\begin{abstract}


In the paper, the model of the society represented by a social network and the model of a multi-agent system built on the basis of this, is presented. The particular aim of the system is to predict the evolution of a society and an analysis of the communities that appear, their characteristic features and reasons for coming into being.
As an example of application, an analysis was made of a social portal which makes it possible to offer and reserve places in rooms for travelling tourists. 

\end{abstract}


%
\IEEEpeerreviewmaketitle

\section{Introduction}

The goal of the work is to analyse and understand rules of social organisation behaviour, activities and evolution directions. 
We are especially focusing our attention on the problem of identifying and evolving the importance of the entities in the social organisation, structure of the formed groups and the reason they emerge. The works carried out are based on methods offered by the social network analysis and the multi-agent approach
\cite{kozlak2010}. 

The application of social simulations makes it possible to analyse processes which occur in society based on behaviour patterns being shaped by numerous simple interactions between the entities \cite{gilbert05}. One kind of social simulation which is especially interesting for us is the social multi--agent--based simulation. This approach takes advantage of agents and useful features of the multi--agent system such as, support of the modelling of a simulated reality, maintaining an organisation structure and a simulation of the proactive behaviour \cite{davidsson00}. 
The goal of the designer of the multi-agent model is to represent a society based on the accessible data about it, including entity features and interactions taking place among them, which make it possible to identify the most important features of the analysed system.  It allows us to analyse the system respecting its dynamic of which one offers a better understanding of activities being performed in the society which, are complex and difficult to interpret. Another task of the described model is to enable the prediction of future situations and future states of the system.
In the paper, the results obtained are also presented thanks to developed pilot applications, which analyse data concerning networking portals and Web blogs.

\section{Overview of research domain}



In this work, two approaches of analysing a society are used -- social network analysis (SNA) and social agent--based simulations (SABS). The essential difference between them is based on the fact that the SNA is based on the analysis of the society as a whole and in SABS the model with the description of local interaction is given and this model is used to design the simulation tool which makes it possible to analyse the society behaviour.


Solutions based on SNA are widely used in both scientific research and practical applications. The social networks are usually considered as a graph where the nodes are usually persons or organisations (actors) and edges are different dependencies between them (phone calls, posts or comments on blogs etc.).


Such analysis may have an important impact while trying to better understand the nature of existing links between the subjects. The basis of such analysis may be - different data with characteristics of given entities, the relations between them and the dynamics of these interactions.


SNA offers methods useful for the analysis of the importance of given entities, roles played by them in the organisations and for the identification of existing strongly-connected communities and their dynamic. SNA offers different measures for the evaluation of characteristics and importance of nodes in the graph such as their degrees of input and output, location on the paths with the highest information flow (Betweeness),  describing connections with important nodes in the graph (Hubness) or with nodes having numerous connections to other important nodes (Authoritativeness), as well as connections with important or well-connected nodes (Page Ranks) etc.


One of the important problems is identifying the communities in a network. Different approaches are used \cite{tang10}, for example  {\it Clique Percolation Method (CPM) Algorithm}, based on identification of k-cliques \cite{palla2008,palla2005}.
For further analysis of the different characteristics describing the communities and their transformation in time \cite{xu2004}  are calculated, which concerns 
the comparison of the strength of internal relations of group members with their external connections with nodes outside the group, density of connections in the group or stability of the membership in time.


An important feature of the organisation analysed by SNA, is the evolution of the organisation in time. The adequate approaches are presented in \cite{mcculloh2008}. One can distinguish different kinds of methods: descriptive, statistic (\cite{snijders2001}) or simulations (\cite{hummon2000}).

Real social networks are highly dynamic: new edges reflecting the appearance of new connections between actors appear or disappear. Kleinberg in \cite{kleinberg2003} called such a process as the {\it link prediction problem}: given information about the state of the network in time {\it t}, we try to predict the edges that will be added to the network to time {\it t'}.
In \cite{kashima2006} two types of information useful for making predictions are recognized about the nodes and network topology. 
In \cite{kleinberg2003, bhagat2010} several predictive models were proposed, for example, Friend-of-a-friend FOAF (for a given node, the  model tries to predict edges to all nodes that are within two hops in the graph; each edge is treated equally), Common-neighbours CN (assumes that there are many common neighbours of two nodes that are more likely to become linked), Adamic-Adar AA \cite{adamic2001} (models assume that all neighbours are not equal and a common neighbour with a low degree is more significant than one with a very high degree), Preferential-Attachment PA \cite{albert2002} (model is based on such properties of the graph that suggests that links are more likely to nodes with a high degree than to ones of low degree). 


To make a deeper exploration of interesting social structures possible, we are going to execute a simulation with the use of the multi--agent approach. This approach is useful because of the simulation of autonomy of given entities.
The goal of the multi--agent based social simulation system is an analysis of the structure of the social organisation, state of the given members of the organisation and  state of communities as well as the directions of the organisation evolution and reasons of these changes. 


The agent approach also offers different models and definitions of measures describing the importance of entities and relations among them. However, in contrary to social networks, which have a global nature of describing the importance of nodes from the point of view of organisation as a whole, the multi-agent approach focuses on local measures describing the reciprocal importance and mutual relations of given pairs of subjects.

\section{Model}



In this section the model of society will be defined, which takes into consideration values of SNA measures for given entities, their domain attributes, roles associated with them as well as the information about interaction between entities and their belongingness to the groups. 
The model presented in section \ref{StaticSocietyModel} is a universal static description of elements used in research as well as fundamental operations. On the basis of information gathered, the model of a multi--agent system is designed, which is used for the analysis of such a society, and especially its evolution. 


\subsection{Static model of the society}
\label{StaticSocietyModel}


The society is represented as follows:
\begin{equation}
Soc= (N, X, C, \upsilon, \eta, \theta, I, R, \psi, \gamma, G, \phi)
\end{equation}
where:


\begin{itemize}
\item $N$ -- set of entities forming a society;
 \item $X$ -- attributes of entities, it contains $X_{D}$ -- domain attributes, $X_{SNA}$ --  attributes containing values of executed SNA measure, $X$  constitutes a sum of all defined attributes for given $j$ entities $X=\bigcup_{j \in N} X^j$;
 \item $C$ -- set of roles, which may be assigned to given entities; 
 \item $\upsilon$ --  assigning of numeric values of attributes for entities: $\upsilon: N \times X \rightarrow \mathbb{R}$; 
 \item $\eta$ -- a function which assigns strengths of association with each role from the set $C$ to every entity  $\eta: N \times C \rightarrow [0,1]$;
 \item $ \theta: N \rightarrow C$ -- a function which assigns a dominant role, it means the one for which the value of $eta$ function is the highest, to every entity;
 \item $I$ --  interactions between entities, which has information about every common influence of the entities, together with the times it took place and their type (for example, in some Internet social network portals, one may have different kinds of links with different values and features of interactions, for example, the exchanged messages, being in the group of friends of a given person, being subscribed to the same groups of interests etc.);
 \item $R$ -- social relations which appear as a result of interactions taking place; 
 \item $\psi$ -- function which establishes social relations $R$ between entities, taking into consideration interactions between entities and their characteristics, $\psi: N \times I \rightarrow R$;
 \item $\gamma$ -- a function which decides the belonging of entities to groups $\gamma: N \times R \rightarrow G$ ;
 \item $G$ -- a set of identified groups; 
 \item $\phi$ -- a function which assigns entities to given groups, specifying the strength of this belonging: $\phi: N \times G \rightarrow [0,1]$
\end{itemize}

\subsection{Description of the entity}

The entities in the society might be described in a similar way. An entity $N^j$  may be described as follows:

\begin{equation}
N^j= (X^j, C, \upsilon^j, \eta^j, \theta^j, I^j, R^j, \psi^j, \gamma^j, G^j, \phi^j)
\end{equation}
The value of arguments concerning only the $j$-th entity $N^j$ are specified by adding the upper index $j$. The functions $\upsilon^j$ $\eta^j$ $\theta^j$, $\psi^j$,  $\gamma^j$,  $\phi^j$ concern only the given entity $N^j$ in the domain, not the whole set of entities $N$
$I^j$ and $R^j$ are interactions and social relations in which the entity $N^j$ participates, $X^j$ -- its set of attributes, $G^j$ -- the groups to which it belongs, set of possible roles $C$ is the same as for the whole society,

\subsection{Dynamic model of the society}

Dynamic model of the society  $Soc(t)$ describes its state in the time $t$ and is expressed as follows: 

\begin{equation}
\begin{split}
Soc(t) = (N(t), X, C, \upsilon_t, \eta_t, \theta_t, \\
I(t), R(t), \gamma_t, G(t), \psi_t, \phi_t)
\end{split}
\end{equation}


where: $N(t)$, $I(t)$, $R(t)$, $G(t)$ values of the adequate sets in the time $t$, and $\upsilon_t$, $\eta_t$, $\theta_t$, $\gamma_t$, $\psi_t$  $\phi_t$ the timely versions of the adequate functions, which may take into consideration the history of changes. 

\subsection{Model of entity evolution}

%

The changes of the entities in the society may concern:
\begin{itemize}
 \item change of the characteristic of the entities; 
 \item change of the set of the entities after joining the new ones.
\end{itemize}

So, the decisions for the modelling are: 
\begin{itemize}
 \item adding a set of new entities $N^{new}$, which arrive in the period [t,t+1], with given characteristics  
        $N(t+1)= N(t) \cup N^{new}$;
 \item set of values of the attributes $X_D^j$ of the entity $N^j$: $X_D^j(t+1)=CX(X_D^j(t))$, where the $CX$ is a function describing the rules of this change.
\end{itemize}


\subsection{Model of social relations and how they evolve}



On the basis of interactions $I$ taking place in the system, it is possible to build a model of social relations between entities $R$. The $\psi$ function which builds the social relation considering interactions between entities is responsible for this task.

Interactions of the entity $N^j$ are represented as $I^j$ and are defined as follows:
\begin{equation}
\begin{split}
 I^j = \{i: i=(N_{in}^i, N_{out}^i, d^i, t^i, c^i), \\
N_{in}^i = N^j\veebar N_{out}^i = N^j\}
\end{split}
\end{equation}
where,
\begin{itemize}
	\item $i$ -- a single interaction;
	\item entities which initiates the interaction ($N_{in}^i$) and  
         receives it $N_{out}^i$;
   \item $d^i$ -- kind of interaction;
 	\item $t^i$ -- time, when the interaction takes place;
   \item $c^i$ -- strength of interaction.
\end{itemize}


The evolution of interactions is described as follows: a set of interactions of the entity  $N^k$ in the time  $t_i$ is represented as  $I^k(t_i)$,  whereas a set of interactions in the time period $\tau_i = [t_{i}, t_{i+1}]$ is represented as $ \hat{I}^k$. 
The evolution of social relations $R^k$ of an entity $N^k$, described below, is represented in a similar way.


The function $\psi$ makes a decision about the existence of social relations and their strengths. 
It is important that these relations represent real and robust relations between entities which emerge from the interactions taking place.
Different kinds of algorithms may be used in determining the relation links. The obtained relation set $R^j$ of the entity $N^j$ contains a set of links with other entities and is represented as follows:
\begin{equation}
\begin{split}
 R^j = \{r^i: r^i=(N_{in}^i, N_{out}^i, s^i, c^i),\\
 N_{in}^i = N^j \veebar N_{out}^i = N^j\}
\end{split}
\end{equation}

where:

\begin{itemize}
	\item $r^i$ -- given relational link;
	\item entities which initiate ($N_{in}^i$) and receive $N_{out}^i$ the relational link;
   \item $s^i$ -- strength of the relational link;
   \item $c^i$ -- semantic of the relational link (for example a model of exchanged content).
\end{itemize}



\subsection{Model of group evolution}


Different methods may be used to determine the memberships of entities in groups/communities. An overview of these is in \cite{tang10}.

A group $G^k$ $\in$ $G$ is described as follows:
\begin{equation}
G^k(t)=(\varmathbb{N^k(t)}, c^k(t))
\end{equation}
where
\begin{itemize}
	\item $\varmathbb{N}^k(t)$ --  a set of entities which belongs to the group in the time $t$;
   \item $c^k(t)$ -- a subject--matter of the group. It may be described for example using a bag of words methods based on every interaction which take place between the members of the group.
\end{itemize}


The main operation which may change the group defined in such a way is when given entities enter or leave the group and an evolution of group subject--matter, which takes place as a consequence of changing the subjects exchanged during the communication within the group.


\subsection{Characteristic of the multi--agent system model}


The goal of the multi--agent system is to develop a model of the society which makes it possible to identify the most important features, such as: 
\begin{itemize}
 \item reasons of group formation and characteristic of groups;
 \item  trends and evolution of entities and groups. 
\end{itemize}


The multi--agent system plays the following roles:
\begin{itemize}
\item makes possible to better understand activities in the systems, which are difficult for surveillance because of their high number and complexity, especially important is taking into consideration the system dynamics during the analysis;
\item	makes it possible to classify a system -- adjust it to given patterns of system functioning; 
\item Predicts future situations and future states of the system. 
\end{itemize}


So, the order of works during the configuration and verification of the multi--agent systems which simulates and analyses organisation is as follows:
\begin{itemize}
\item gathering characteristics describing the state of the system, entities and groups as well as its evolution in the subsequent time periods $\tau_1, \ldots, \tau_N$;
\item configuring the multi-agent system according to the gathered characteristics  and its launching; 
\item comparing results delivered by the multi--agent system with the results obtained during the analysis of data using SNA, modification of algorithms used in the multi--agent system to provide a better reflection of the activities in reality, this step may be repeated subsequently for different selected  observation time periods until the expected quality of the results is obtained;
\item launching the configured multi--agent system for predicting the future states of the society.
\end{itemize}



The multi--agent system needs the following information and makes it possible to predict the changes of the following elements:
\begin{itemize}
\item set of attributes of the agents describing the motivation of its activities;
\item information about the social roles assigned to the agents; 
\item interactions of the given types among the agents in the given time periods and the subsequent periods; 
\item social relations between the agents in the given time periods and subsequent time periods as well as their strengths;
\end{itemize}


\subsection{Decision functions for agents}

For designing such a system it is necessary to prepare necessary decision functions which describe the activities to be performed by the agents.
Analysis of the agent activities are based especially on the updating of the state of attributes as well as performing  interactions of adequate types
with adequate partners and building social relationships in consequence. On their basis, agents may have assigned roles as well as their memberships in groups being determined. The methods similar to the ones presented in the society description might be applied.



The decision function $CX$ which describes the evolution of the values of attributes in time is defined as follows:
\begin{equation}
X^j(t+1)= CX(X^j(t), \upsilon^j, I^j(t), R^j(t))
\end{equation}
where:

\begin{description}
 \item $X^j$ -- attributes of agent $j$; 
 \item $\upsilon^j$ --  functions which assigns values to attributes;
 \item $I^j$-- interactions of agent $j$;
 \item $R^j$ -- social relations of agent $j$;
\end{description}


The next two decision functions describe the decisions concerning the performed interactions.
Function $CWI$ (calculate willingness to interactions) decides which new interaction should take place during the next time period. It is defined as follows: 
\begin{equation}
WI^j(t+1)= CWI (X^j(t), \upsilon^j, I^j(t), R^j(t))
\end{equation}
where $WI^j(t+1)$ represents the willingness to initiate the interactions 
and function arguments are similar to the arguments of $CX$ function.


Some interactions need not only the decision of their initiator to take place, but also the acceptance of the interaction receiver. For example, in the social networking portals for establishing a link which represents acquaintance or friendship the receiver of the proposition may accept or reject it.
Similar situations takes place in phone calls, the receiver has to take the call.

In such situations the decision is defined by the $CWAI$ (calculate willingness for interaction acceptance function described as follows:
\begin{equation}
WAI^j(t+1)= CWAI(X^j(t), \upsilon^j, I^j(t), R^j(t))
\end{equation}
where $WAI$ represents willingness for interaction acceptance and function arguments are similar to the ones used in previous functions.


The function $CR$ (calculate relations) calculates the evaluation of social relations of the agent with other agents and is defined as follows:
\begin{equation}
\begin{split}
R(t+1) = CR(R^j(t+1), X^j(t+1),\upsilon^j,\\
WI(t+1), WAI(t+1))
\end{split}
\end{equation}
where:
\begin{description}
 \item $R$ -- current value of social relations in time;
 \item $X$ -- attributes,
  \item $\upsilon^j$ --  functions which assign values to attributes;
 \item $WI$ --  willingness for interactions;
 \item $WAI$ -- willingness for interaction acceptance.
\end{description}

\subsection{Models of agent and multi--agent system}

The multi--agent system $MAS$ in time $t$ is defined as follows:
\begin{equation}
MAS(t) = (IA, \varmathbb{A(t)})
\end{equation}
where

\begin{itemize}
 \item $IA$ -- criteria describing entering agents with given characteristics into the system;
 \item {\it A}  -- set of agents in the systems which includes agents of $AN^j$ and $AG^k$ types.
\end{itemize}

There are two kinds of agents in the system:
\begin{itemize}
 \item agents which represent entities/users ($AN^j$),
 \item agents which represent groups of users ($AG^k$).
\end{itemize}

Each agent $AN^j$ represents an entity $N^j$ in the system and is an autonomous element modelled on the basis of information gathered by the entity $N^j$, it may be represented as follows: 
\begin{equation}
\begin{split}
 AN^j = (X^j, C^j,  \upsilon_t^j, \eta_t^j, \theta_t^j, I^j(t), R^j(t), \phi_t^j, \\
CX, CWI, CWAI, CR)
\end{split}
\end{equation}

The agent representing $AG^k$ represents groups $G^k$ are described as follows:
\begin{equation}
\begin{split}
AG^k=\{G^k, \upsilon_t^k, \eta_t^k, \theta_t^k, I^k(t), R^k(t),\\
CX, CWI, CWAI, CR\}
\end{split}
\end{equation}

The parameters for both entity agents and group agents are as follows:

\begin{itemize}
	\item $G^k$ -- the group defined previously which is represented by $AG^k$ agent;  
   \item $\upsilon_t^k$ --  assigning of the attributes values; 
   \item $\eta_t^k$ -- assigning of the roles evaluation; 
   \item $\theta_t^k$ -- assigning of the dominant roles; 
   \item $I^k(t)$ --  interactions with other agents and groups;
   \item $R^k(t)$ -- relationships with other agents and groups; 
   \item $CX$ --  function which calculates a new state of the agent,
   \item $CWI$ -- function which calculates a willingness for interaction for the agent; 
   \item $CWAI$ -- function which calculates a willingness for interaction acceptance for the agent.
   \item $CR$ -- calculate new relations,
\end{itemize}

%
%
%

\section{Realisation}

%
%


As a subject of research we selected the portal couchsurfing.net, which is used by the society of people from different countries who are often travelling. They present information about their travels.
The most important role of the portal is to facilitate finding a place to stay overnight during these travels. Users offer places in their houses or apartments or ask for places to stay in the cities they want to visit.

For this portal we performed an analysis using the methodology of SNA. On the basis of gathered information, the multi--agent system was designed, which makes it possible to predict future behaviour of users, considering the established connections and arrival of new users with given characteristics.

For assigning roles of the following domain, attributes were introduced:
\begin{itemize}
  \item {\it friendsmaker} –-  activity of establishing contacts with new people; 
  \item {\it talker} –- activity in thematic groups;
  \item surfer –-  activity in using the portal for finding lodging for the nights during travels;
  \item {\it host} –- activity in offering accommodation;
  \item {\it traveller} –- activity in finding travel mates.
\end{itemize}


In the research, also two classical SNA measures describing the position of nodes in the graph were used: closeness (a distance in the network needed for propagation of information, expressed as a reverse sum of the shortest paths from a given node to every other node in the graph) and betweeness (which describes the importance of the given node as a bridge between groups.)

%


On the basis of these parameters, the users of the system were assigned to the following groups:
\begin{itemize}
 \item {\it Host} – a person who offers night accommodation for other users, he/she also travels frequently;
 \item {\it Traveller} – a person who often travels and also offers accommodation/lodging to other users, it is similar to Host, but they travel more than the host;
\item {\it Virtual} –- a person who has many acquaintances and makes journeys and houses users;
\item {\it Homebody} –- a person who often houses other users, but his/her travelling activity is very low;
\item  {\it Scrounger} –- an opposite of Homebody, often takes advantage of hospitality of others, but offers their own accommodation very rarely;
\item {\it Observer} –- a person who owns an account on the portal but does not show any activity among the presented above or shows it to a very low degree. 
\end{itemize}



\begin{figure}[!t]
\centering
\includegraphics[width=2.5in]{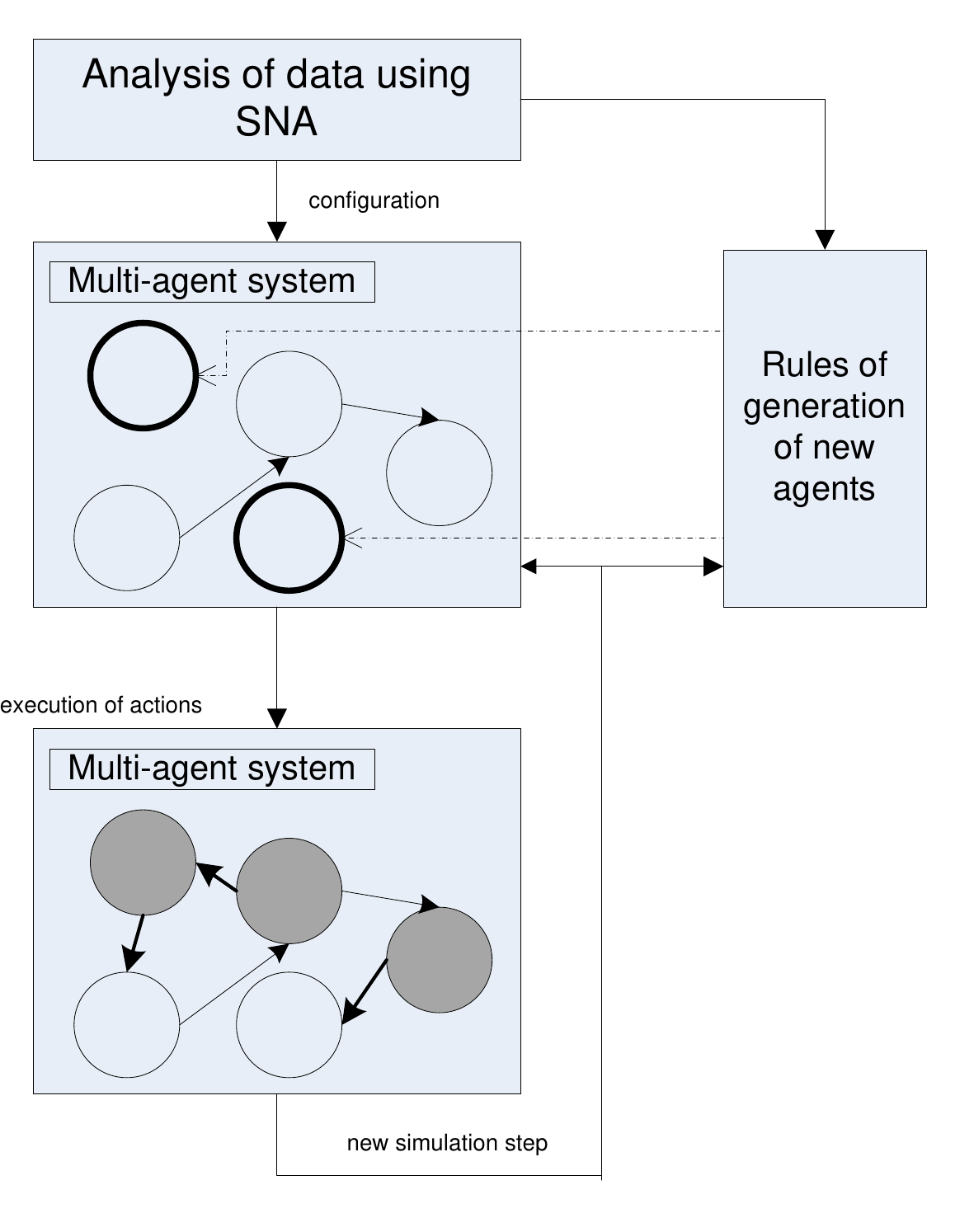}
\caption{General schema of simulator functioning}
\label{fig:alg}
\end{figure}

{\it Algorithm description}. 
The general schema of the algorithm is presenten in fig. \ref{fig:alg}.

At the beginning the initial population of agents representing users is generated. These agents are configured according to the results and value distributions obtained using social network analysis. The following actions of the algorithm are repeated many times and are as follows:
\begin{itemize}
 \item a generation of the adequate number of new agents for this step;
 \item assigning of roles and profiles to them;
 \item execution of actions by every agent in the population. The agent behaviour is determined by two main factors: attributes and functions which determine a probability of performing given actions. In each step, the following actions might be performed:
\begin{itemize}
  \item establishing of the acquaintance;
  \item decision concerning a common travel; 
  \item decision concerning lodging a user; 
  \item decision concerning visiting a user;
\end{itemize}
\end{itemize}

\section{Results}

%

Below, the selected experimental scenarios carried out using a pilot application are presented
\begin{itemize}
 \item analysis using SNA  -- 
the values of SNA and domain measures were calculated and roles from the defined set were assigned to users. The comparison of user behaviour in depending on their declared country of origin was done so as to observe the specificity of the behaviour of people with different social backgrounds.

 \item analysis using the multi--agent system -- 
the multi--agent system was configured on the basis of activities coming from the initial period of social network behaviour and then a simulation of network evolution was performed. The analysis especially concerned the prediction of assigned roles and the appearance of new nodes in the network.

 \item evaluation of the prediction quality -- the system characteristics obtained using the social network analysis were compared with the results of agent--based simulator. 
\end{itemize}


\begin{figure}[!t]
\centering
\includegraphics[width=2.5in]{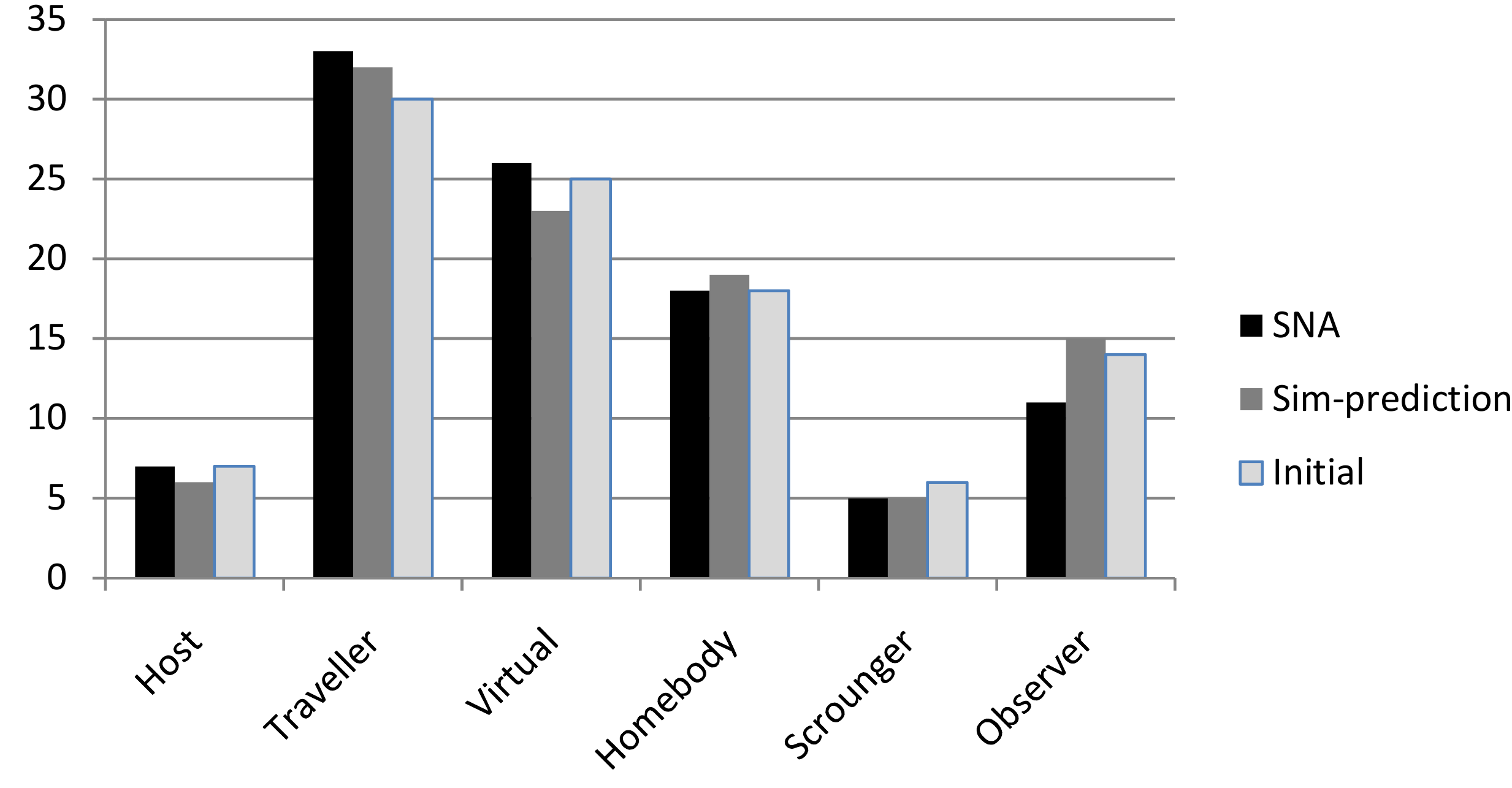}
\caption{Distribution of roles in percentages. SNA - results of social network analysis, Sim-prediction -- results of simulation, Initial - distribution for initial perdiod used for the configuration of the MAS simulation}
\label{fig:sim1}
\end{figure}

\begin{figure}[!t]
\centering
\includegraphics[width=2.5in]{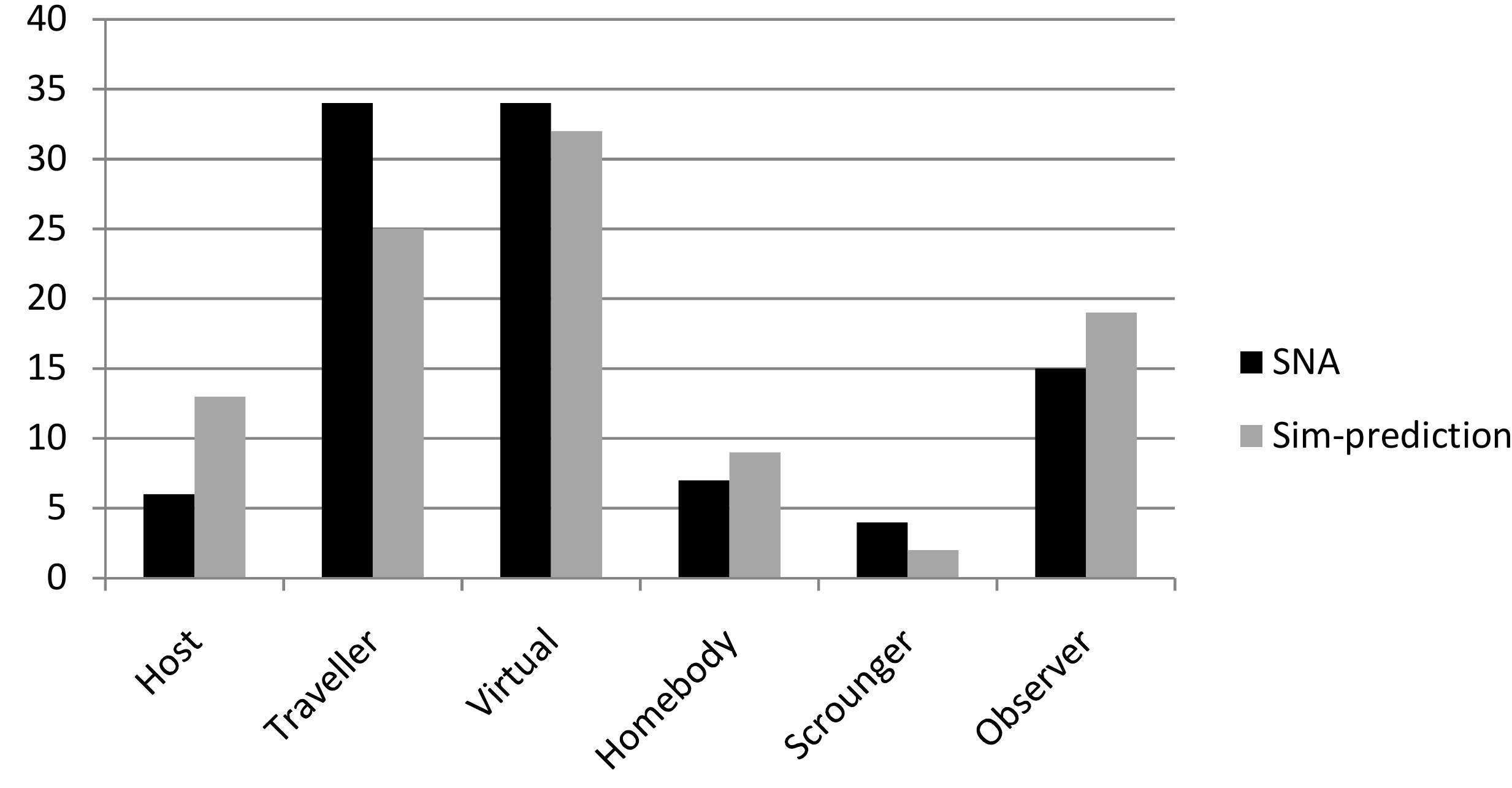}
\caption{Distribution of roles in percentages for entities added after the initial period. SNA - results of social network analysis, Sim-prediction -- results of simulation.}
\label{fig:sim2}
\end{figure}


{\it Analysis used social networks approach.}
For the analysis of data about users from 12 countries in Central--Eastern Europe. The roles assigned to users depending on their country of origin were analysed. The results show a stronger use of the portal for touristic goals by the users from wealthier countries. In the cases of poorer countries, the roles not associated with intensive travelling were more often assigned.


{\it Comparison of percentage of roles assigned by SNA and MAS}. The results of the prediction obtained by MAS simulator seems to be satisfactory (fig. \ref{fig:sim1}). For the majority of roles, differences are not very high. 
The highest variations take place for roles of Traveller and Observer -- a higher number of Observers and lower number of Travellers were predicted than were actually calculated in reality. It may be explained by time periods when the data was gathered.  The initial data, which was the basis for the configuration of the simulator concerns the time periods before summer holidays while the important part of the rest of the data was gathered during this holiday period.
One can assume that during this period the number of trips increased and some Observers change the behaviour to Travellers which was not taken into consideration during the simulation.


{\it Comparison of the percentages of assigned new roles.}
The difference of roles assigned to new users obtained by SNA and by MAS prediction was higher than in the case of the whole population (fig. \ref{fig:sim2}). This may be a consequence of a higher random disturbance as well as a smaller sample used for analysis. 

\section{Conclusion}




In the paper, the universal model of the society which was changing in time and the multi-agent model for modelling the selected aspects of its behaviour were presented. 
Different functions for data analysis were presented. Depending on the form of the function, different kinds of the multi--agent model might be obtained with different relations between entities and different characteristics of agents and groups.

\section*{Acknowledgment}

The authors would like to thank students and former students of Computer Science on Department of Computer Science AGH-UST, especially 
\L{}ukasz Grzyb and Micha\l{} Grad who participated in the implementation of the systems.



\bibliographystyle{IEEEtran}
\bibliography{bib}

\begin{thebibliography}{10}
\providecommand{\url}[1]{#1}
\csname url@samestyle\endcsname
\providecommand{\newblock}{\relax}
\providecommand{\bibinfo}[2]{#2}
\providecommand{\BIBentrySTDinterwordspacing}{\spaceskip=0pt\relax}
\providecommand{\BIBentryALTinterwordstretchfactor}{4}
\providecommand{\BIBentryALTinterwordspacing}{\spaceskip=\fontdimen2\font plus
\BIBentryALTinterwordstretchfactor\fontdimen3\font minus
  \fontdimen4\font\relax}
\providecommand{\BIBforeignlanguage}[2]{{%
\expandafter\ifx\csname l@#1\endcsname\relax
\typeout{** WARNING: IEEEtran.bst: No hyphenation pattern has been}%
\typeout{** loaded for the language `#1'. Using the pattern for}%
\typeout{** the default language instead.}%
\else
\language=\csname l@#1\endcsname
\fi
#2}}
\providecommand{\BIBdecl}{\relax}
\BIBdecl

\bibitem{gilbert05}
N.~Gilbert and K.~G. Troitzsch, \emph{Simulation for the the Social
  Scientist}.\hskip 1em plus 0.5em minus 0.4em\relax Open University Press,
  2005.

\bibitem{davidsson00}
P.~Davidsson, ``Multi agent based simulation: Beyond social simulation,'' in
  \emph{MABS}, 2000, pp. 97--107.

\bibitem{tang10}
L.~Tang and H.~Liu, \emph{Community Detection and Mining in Social
  Media}.\hskip 1em plus 0.5em minus 0.4em\relax Morgan \& Claypool Publishers,
  2010.

\bibitem{palla2008}
G.~Palla, D.~Ábel, I.~J. Farkas, P.~Pollner, I.~Derényi, and T.~Vicsek:,
  ``k-clique percolation and clustering,'' in \emph{Handbook of Large-scale
  Random Networks}, B.~Bollobás, R.~Kozma, and D.~Miklós, Eds.\hskip 1em plus
  0.5em minus 0.4em\relax Springer, 2009.

\bibitem{palla2005}
G.~Palla, I.~Derenyi, I.~Farkas, and T.~Vicsek, ``Uncovering the overlapping
  community structure of complex networks in nature and society,''
  \emph{Nature}, vol. 435, pp. 814--818, 2005.

\bibitem{xu2004}
J.~Xu, B.~Marshall, S.~Kaza, and H.~Chen, ``Analyzing and visualizing criminal
  network dynamics: A case study,'' in \emph{IEEE Conference on Intelligence
  and Security Informatics}, Tuczon, 2004.

\bibitem{mcculloh2008}
I.~A. McCulloh and K.~M. Carley, ``Social network change detection,'' Carnegie
  Mellon, San Francisco, Tech. Rep. CMU-CS-08-116, March 2008.

\bibitem{snijders2001}
T.~Snijders, ``The statistical evaluation of social network dynamics,'' in
  \emph{Sociological Methodology Dynamics}, M.~Sobel and M.~Becker, Eds.\hskip
  1em plus 0.5em minus 0.4em\relax London: Basil Blackwell, 2001, pp. 361--395.

\bibitem{hummon2000}
N.~Hummon, ``Utility and dynamic social networks,'' \emph{Social Networks},
  vol.~22, no.~3, pp. 221--249, 2000.

\bibitem{kleinberg2003}
D.~Liben-Nowell and J.~Kleinberg, ``The link prediction problem for social
  networks,'' in \emph{Proc. 12th International Conference on Information and
  Knowledge Management (CIKM)}, 2003.

\bibitem{kashima2006}
H.~Kashima and N.~Abe, ``A parameterized probabilistic model of network
  evolution for supervised link prediction,'' in \emph{Proc. of the Sixth
  International Conference on Data Mining ICDM'06}, 2006.

\bibitem{bhagat2010}
S.~Bhagat, G.~Cormode, B.~Krishnamurthy, and D.~Srivastava, ``Prediction
  promotes privacy in dynamic social networks,'' in \emph{Proceedings of the
  3rd conference on Online social networks (WOSN'10)}, 2010.

\bibitem{adamic2001}
L.~Adamic and E.~Adar, ``Friends and neighbors on the web,'' \emph{Social
  Neworks}, vol.~25, 2001.

\bibitem{albert2002}
R.~Albert and A.-L. Barabasi, ``Statistical mechanics of complex networks,''
  \emph{Rev. Mod. Phys.}, vol.~74, 2002.

\end{thebibliography}
%



\end{document}